\documentstyle[multicol,aps,prl,epsf]{revtex}

\begin{document}

\draft

\title{
A link between the spin fluctuation and Fermi surface in high $T_C$ cuprates\\
--- A consistent description within the single-band Hubbard model
}

\author{
Kazuhiko Kuroki, Ryotaro Arita, and Hideo Aoki
}
\address{Department of Physics, University of Tokyo, Hongo,
Tokyo 113-0033, Japan}

\date{\today}

\maketitle

\begin{abstract}
A link between the spin fluctuation and the ``fermiology" 
is explored for the single-band Hubbard model 
within the fluctuation exchange (FLEX) approximation .  
We show that the experimentally observed peak position of the spin 
structure in the high $T_C$ cuprates can be understood from 
the model that reproduces the experimentally observed Fermi surface. 
In particular, both the variation of the incommensurability 
of the peak in the spin structure 
and the evolution of the Fermi surface with hole doping
in La$_{2-x}$Sr$_x$CuO$_4$ may be understood 
with a second nearest neighbor hopping decreasing with hole doping.
\end{abstract}

\medskip

\pacs{PACS numbers: 71.10.Fd, 74.25.Ha, 74.72.-h }

\begin{multicols}{2}
\narrowtext

\newpage
In the physics of high $T_C$ cuprates, it is of 
great importance to investigate how the states 
dominated by the electron-electron interaction are affected by
the band structure and the band filling.
This is especially so for the spin fluctuation, which 
has been recognized as one of the central normal state 
properties.  So it is an important challenge to look into 
the link between the spin fluctuation and the ``fermiology".  

The doping dependence of the spin fluctuation 
in La$_{2-x}$Sr$_x$CuO$_4$ (LSCO) has been well established from 
neutron scattering experiments, where incommensurate spin structures appear 
when holes are doped.\cite{review}
It has also been shown that such an incommensurate spin structure can
be understood qualitatively within the single-band Hubbard or $t$-$J$ models.
\cite{Tremblay,Si,TKF,Duffy,Avella,Moreo,Kontani}
However, the doping dependence of 
the incommensurability $\delta$ (the deviation of the 
peak in the spin structure from ${\bf q} = (\pi,\pi)$) 
has not been, to our knowledge, 
understood quantitatively within single-band models. 
Namely, large $\delta$ appears with a small amount of hole doping 
(e.g., $\delta\simeq 0.2$ already at $x\sim 0.1$) in the 
underdoped regime, while $\delta$ saturates to $0.25 \sim 0.30$ 
in the overdoped regime.\cite{Yamada1} 

More recently, the incommensurate spin structure is observed in
the (underdoped) YBa$_2$Cu$_3$O$_{6.6}$ (YBCO$_{6.6}$) and 
(nearly optimally doped) Bi$_2$Sr$_2$CaCu$_2$O$_8$ (Bi2212) samples 
as well,\cite{Dai,Mook1,Mook2} 
which suggest that the phenomenon is not an accident for LSCO.
Several authors have theoretically investigated 
the Hubbard\cite{Avella,Moreo} and $t$-$J$ models\cite{Lee} 
in the context of these materials.  
By contrast, it is well known that the antiferromagnetic order
is robust against electron doping in Nd$_{2-x}$Ce$_x$CuO$_{4-y}$ (NCCO).
The difference is endorsed by a recent observation of 
a commensurate spin fluctuation 
in a superconducting NCCO sample $(x=0.15)$.\cite{Yamada2}

On the other hand, the shape of the Fermi surface 
has been revealed from the angle resolved photoemission (ARPES) studies 
for various high $T_C$ 
cuprates including YBCO\cite{Campuzano}, Bi2212\cite{Dessau,Ding},
LSCO\cite{Fujimori,Ino}, and NCCO\cite{King}.
In particular, a recent ARPES experiment\cite{Fujimori,Ino} has
shown that the nesting of the 
Fermi surface in the underdoped LSCO is not so good as has been expected 
previously\cite{Hybertsen1,Radtke} from band 
calculations.\cite{Freeman,Hybertsen2}

So the theoretical challenge is the following.  
We can simulate these shapes of the Fermi surface by 
taking appropriate values for the hopping parameters 
extended to distant neighbors in the tight-binding model, 
and the question is how 
the electron-electron interaction would dictate the spin structure. 
This is exactly the purpose of the present study.  
We employ here the single-band Hubbard 
model, one of the simplest models for repulsively interacting electrons.
We shall show that the experimentally observed Fermi surface and 
the peak position of the spin structure can be understood consistently 
within the single 
band Hubbard model by taking appropriate values for the hopping parameters.
The spin susceptibility and 
the Fermi surface of the Hubbard model 
are calculated with the fluctuation exchange (FLEX) 
approximation,\cite{Bickers} 
which is appropriate for treating large spin fluctuations. 

We consider the single band Hubbard model,
\begin{equation}
H=-\sum_{\langle i,j\rangle,\sigma}
t_{ij}(c^\dagger_{i\sigma}c_{j\sigma}+{\rm H.c})
+U\sum_i n_{i\uparrow}n_{i\downarrow},
\end{equation}
where we take $t_{ij}=t$ for nearest neighbors, $t_{ij}=t'$ for 
second nearest neighbors, and $t_{ij}=t''$ for third nearest neighbors.
Using FLEX approximation, we calculate the RPA-type spin susceptibility 
$\chi({\bf q},\omega)$ as well as the Fermi surface.  
In the actual calculation, we take $64\times 64$ $k$-point meshes and 
the Matsubara frequencies $\omega_n$ from 
$-(2N_c-1)\pi T$ to $(2N_c-1)\pi T$, where $N_c=1024$. 
The dynamical spin susceptibility for $\omega\neq 0$ 
is obtained by an analytical continuation of $\chi({\bf q},i\omega_n)$
using Pad\'{e} approximation.\cite{Vidberg}
For all the case studied, we take $U=4t$ and $T=0.05t$.
We denote the band filling as $n$, 
the hole doping level as $p_h=1-n$, 
and the electron doping level as $p_e=n-1$.

\begin{figure}
\begin{center}
\leavevmode\epsfysize=40mm \epsfbox{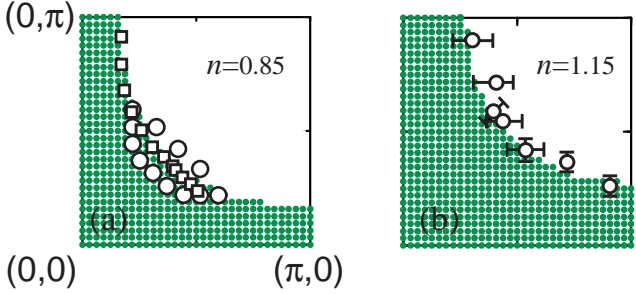}
\caption{The Fermi surface 
for $n=0.85$ (a) or $n=1.15$ (b) 
with $U=4t$, $t'=-0.40t$, $t''=0.05t$, $T=0.05t$ is plotted by gray circles. 
The experimentally observed 
Fermi surfaces of YBCO ($\bigcirc$ (a))\protect\cite{Campuzano}, 
Bi2212 ($\Box$ in (a))\protect\cite{Ding}, 
and NCCO ($\bigcirc$ in (b))\protect\cite{King} are 
superposed for comparison.}
\label{fs-ybco-ncco}
\end{center}
\end{figure}

First let us look at the case of YBCO, Bi2212, and NCCO.
We find that the Fermi surface obtained by ARPES experiments 
for these materials can be reproduced by 
fitting $t'=-0.40$, $t''=+0.05$.
These values are close to the ones adopted in some of the previous studies.
\cite{Duffy,Avella,Lee,Radtke,Tohyama}
In Fig.\ref{fs-ybco-ncco}, we superpose the experimentally obtained 
Fermi surface to that of the Hubbard model obtained with FLEX. 
The fit for YBCO is not so good as the others
because the Fermi surface splits due to the interlayer coupling.

\begin{figure}
\begin{center}
\leavevmode\epsfysize=35mm \epsfbox{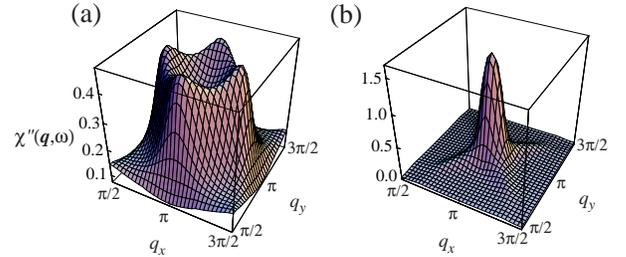}
\caption{$\chi''({\bf q},\omega=0.05t)$ 
plotted against {\bf q} for $n=0.90$ (a) or $n=1.15$ (b) 
with $U=4t$, $t'=-0.40t$, $t''=0.05t$, and $T=0.05t$.}
\label{sus-ybco-ncco}
\end{center}
\end{figure}

Fixing $(t',t'')$ at $(-0.40,0.05)$, we have calculated the 
spin susceptibility $\chi({\bf q},\omega)$ for various band fillings.
In Fig.\ref{sus-ybco-ncco}, 
we show typical results for the imaginary part 
($\equiv \chi''$) of $\chi({\bf q},\omega)$ with $\omega=0.05t$ 
for hole-doped (a) or electron-doped (b) cases.  
We can see that incommensurate
peaks appear at ${\bf q}=(\pi,\pi(1\pm\delta))$, $(\pi(1\pm\delta),\pi)$ for 
hole doping, while a commensurate peak appears 
at ${\bf q}=(\pi,\pi)$ for the electron doping.

\begin{figure}
\begin{center}
\leavevmode\epsfysize=55mm \epsfbox{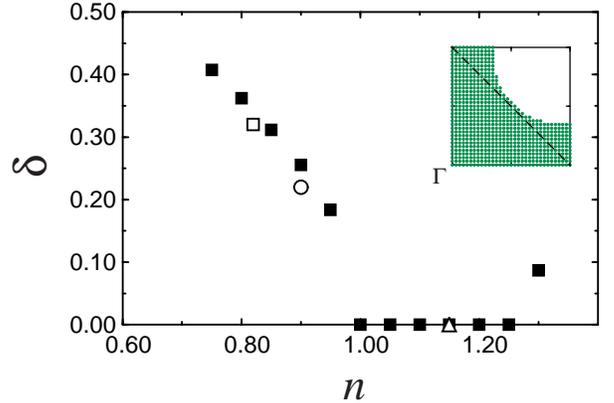}
\caption{Incommensurability $\delta$ plotted as a function of band filling
$n$ for $U=4t$, $t'=-0.40t$, $t''=0.05t$, $T=0.05t$ (solid squares). 
We plot here $\delta$ defined for $\chi({\bf q},\omega=0)$, 
which coincides with $\delta$ defined for $\chi''({\bf q},\omega\neq 0)$ 
within 0.02 for the cases studied. Also plotted are the 
experimental results for YBCO$_{6.6}$ ($\bigcirc$), Bi2212($\Box$), 
and Nd$_{1.85}$Ce$_{0.15}$CuO$_{4-y}$ ($\triangle$).
The inset shows the Fermi surface for $n=1.30$.}
\label{incm-ybco-ncco}
\end{center}
\end{figure}

The asymmetry between the hole and electron doping 
becomes even clearer if we plot the incommensurability $\delta$ 
as a function of the band filling\cite{comment1}
in Fig.\ref{incm-ybco-ncco}. 
One can immediately notice that a rather large incommensurability appears with 
a small amount of hole doping ($\delta\sim 0.25$  at $p_h\sim 0.1$), 
while the commensurate spin structure
is robust against the electron doping up to $p_e \simeq 0.25$.
The robustness of the commensurate spin correlation 
in the electron doped regime is consistent with previous numerical 
studies for the Hubbard model\cite{Duffy} and the $t$-$J$ model.
\cite{Tohyama,Gooding}  
We have further found that 
the obtained onset of the incommensurate spin structure 
around $n \sim 1.3 (p_e \sim 0.3)$ 
in the electron doped regime coincides with 
the situation at which the Fermi surface becomes too bloated
to intersect the magnetic Brillouin zone boundary 
(the line $|k_x|+|k_y|=\pi$), 
as shown in the inset of Fig.\ref{incm-ybco-ncco}.\cite{comment2}

The result agrees quantitatively with 
experimental results for YBCO$_{6.6}$, Bi2212, and NCCO, 
plotted as open symbols in Fig.\ref{incm-ybco-ncco}.  
Namely, YBCO$_{6.6}$, in which the hole doping level in the CuO$_2$ planes 
is around $p_h\sim 0.1$,\cite{Tallon} is shown to have 
$\delta\simeq 0.22$.\cite{Mook1,Mook2} For Bi2212, a neutron
\noindent
scattering result has been obtained for a nearly 
optimally doped sample, in which an incommensurate spin structure has 
been detected. Although the actual spatial pattern of the 
spin structure has not been determined precisely, 
it has been concluded in Ref.\onlinecite{Mook2} that 
$\delta$ would be around 0.32 if the spatial symmetry of 
the spin structure is the same as that in YBCO.
The optimal hole concentration for Bi2212 is around $p_h\sim 0.18$
according to Ref.\onlinecite{Loeser}, which is adopted in the plot. 
For NCCO, the commensurate antiferromagnetic order persists up to 
about $x\simeq 0.13$. Quite recently, a commensurate spin fluctuation
has been found in the superconducting samples 
Nd$_{1.85}$Ce$_{0.15}$CuO$_{4-y}$.\cite{Yamada2}
Here we plot this result by nominally taking $n=1.15$, 
although the actual doping level may depend on $y$. 

A few comments are in order at this point. 
Some previous studies have adopted a smaller 
$-t'(=0.17t)$ and a larger $t''(=0.20t)$for YBCO.\cite{TKF,Kontani}
Although this parameter set approximately reproduces 
the experimentally obtained Fermi surface as well,\cite{TKF,Kontani} 
it only gives a broad commensurate spin structure 
in the underdoped regime ($p_h\leq 0.1$), which crosses over to an 
incommensurate structure with peaks 
in the {\it diagonal} direction ($(\pi(1\pm\delta),\pi(1\pm\delta))$)
rather than for $(\pi,\pi(1\pm\delta))$ 
in the overdoped regime ($p_h\geq 0.2$).

In fact, we can reduce the value of $-t'$ without much changing
the shape of the Fermi surface by taking larger values of $t''$, 
but with a slight reduction of $-t'$ to $t'=-0.30t$ (with $t''=0.10t$),
the incommensurate spin structure at $n=0.90$, although still present, 
is already substantially degraded as compared with that for $t'=-0.40t$. 
Thus, among the sets of parameters that reproduce the Fermi surface of YBCO,
the ones with $-t'\gg t''$ seem to be 
appropriate from the viewpoint of the spin structure.\cite{comment3}

In Ref.\onlinecite{Moreo}, a relatively small $-t'(=0.25t$ with $t''=0$) 
has been concluded on the ground that 
a quantum Monte Carlo calculation shows an appearance of the peak 
in the spin structure in the diagonal direction for $-t'\geq 0.3t$.  
We have also found that the diagonal peaks can appear
for the present choice of $-t'=0.40t$ (with $t''=+0.05t$), 
but that occurs only for band fillings $n=0.70$ (for which the calculation 
in Ref.\onlinecite{Moreo} was performed) or less, 
so that there is no inconsistency. 
However, a smaller $-t'$ is not acceptable in our view, 
since, not only the large curvature of the Fermi surface 
requires a large $-t'$, but 
also the appearance of large $\delta$
with a small amount of hole doping requires a large $-t'$ as we shall see 
in Fig.\ref{incm-lsco} below.  
Since $\delta$ is as large as $\sim 0.22$
for YBCO$_{6.6}$ (with $p_h$ being as small as 0.1), 
a large value of $-t'$ such as the one adopted here is 
considered to be appropriate for this material. 

\begin{figure}
\begin{center}
\leavevmode\epsfysize=70mm \epsfbox{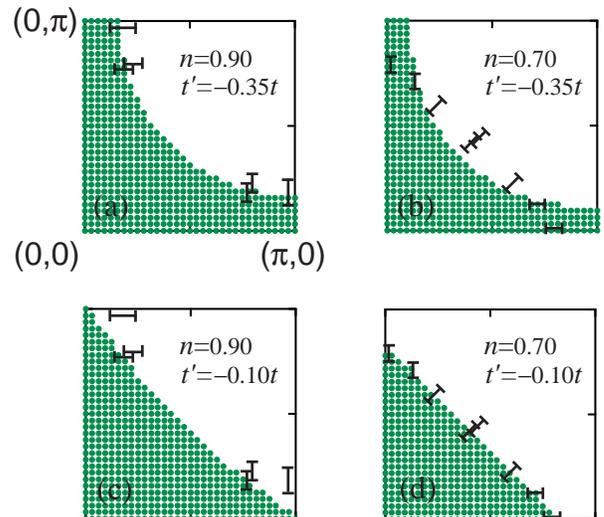}
\caption{The Fermi surface for 
$U=4t$, $T=0.05t$, $t''=0.05t$, and $(t',n)=(-0.35t,0.90)$ (a),
$(-0.35t,0.70)$ (b), $(-0.10t,0.90)$ (c), 
$(-0.10t,0.70)$ (d). The experimentally observed Fermi surfaces of
LSCO for $x=0.1$ (a,c) and $x=0.3$ (b,d)\protect\cite{Ino} 
are superposed for comparison.}
\label{fs-lsco}
\end{center}
\end{figure}

Let us now turn to La$_{2-x}$Sr$_x$CuO$_4$, which 
provides an excellent test-bed for studying the link between the 
Fermi surface and spin structure, since (i) 
the material is unique in that the Fermi surface changes its 
topology (i.e., connectivity) between the underdoped ($x=0.1$) and 
overdoped ($x=0.3$) regimes, as has been observed by a recent ARPES experiment,
\cite{Fujimori,Ino} (ii) the overall doping dependence of the spin structure 
is well established from neutron scattering experiments.\cite{Yamada1}
First let us look into the Fermi surface.
The ARPES result\cite{Fujimori,Ino} suggests that the nesting 
of the Fermi surface for $x=0.1$ is not so good as has been expected 
previously from band calculations, 
so one needs to take relatively large values 
for $-t'$ to reproduce that.  
We can then take $t'=-0.35t$, $t''=+0.05t$, and $n=0.90$ to 
obtain a good fit to the experimentally obtained 
Fermi surface as shown in Fig.\ref{fs-lsco}(a).
On the other hand, the ARPES data at $x=0.3$ 
cannot be reproduced, as seen in Fig.\ref{fs-lsco}(b), 
for fixed $(t', t'') (=(-0.35t, +0.05t)$) with $n=0.70$.  
In fact, we have to assume $-t'$ to be as small as $0.10t$ 
for a fixed $t''(=0.05t)$, or 
$t'=-0.15t$ when $t''$ is varied to zero, to reproduce the ARPES data. 
These values are actually close to
the ones adopted for LSCO in many of the previous studies.
\cite{Tremblay,Si,TKF,Duffy,Avella,Kontani,Hybertsen1,Radtke} 
We show in Fig.\ref{fs-lsco}(d)
how the Fermi surface is well reproduced 
for $t'=-0.10t$, $t''=0.05t$ with $n=0.70$.  
A change in $t'$ is really required, since if we 
fix $(t',t'')$ to these values to go back to $n=0.90$ 
in Fig.\ref{fs-lsco}(c), the Fermi surface is seen to deviate from 
the ARPES data for $x=0.1$.

One might suspect that 
the reason why we have to take different values for the 
hopping parameters in the underdoped and overdoped regimes
might be because the FLEX approximation did not take into 
account properly the electron correlation effect, which should become
more significant in the underdoped regime.
This, however, cannot be the case, since it is known that 
the electron correlation effect tends to {\it enhance} the nesting in 
Fermi surfaces as the system approaches 
half-filling, while the experimental data suggest 
that the nesting becomes {\it worse} for smaller $x$.
Thus we can assume that the decrease of $-t'/t$ with hole doping 
is not an artefact.

Keeping the above analysis for the Fermi surface in mind, we 
have calculated $\chi({\bf q},\omega)$ for three values of $t'$, namely,
$t'/t=-0.35$, $-0.25$, and $-0.10$, fixing $t''=0.05t$.
A typical result for $t'=-0.25t$, $t''=+0.05t$, and $n=0.80$ 
is given in the inset of Fig.\ref{incm-lsco}. Again, incommensurate peaks
are found at ${\bf k}=(\pi,\pi(1\pm\delta))$, $(\pi(1\pm\delta),\pi)$.
The band filling dependence of the incommensurability is plotted in
Fig.\ref{incm-lsco}.
Here, a result similar to that for $(t',t'')=(-0.10t,0.05t)$ can also be 
obtained for $(t',t'')=(-0.15t,0.0)$, i.e., 
the other choice of parameter set that reproduces the ARPES data for $x=0.3$.
The doping dependence of $\delta$ for small values of $-t'$ is similar to those
obtained in previous theoretical studies for LSCO.\cite{Tremblay,Duffy} 
As already stated previously,\cite{Tremblay,Duffy} $\delta$ is 
essentially zero in the underdoped regime
($n\geq 0.90$) for 
\begin{figure}
\begin{center}
\leavevmode\epsfysize=60mm \epsfbox{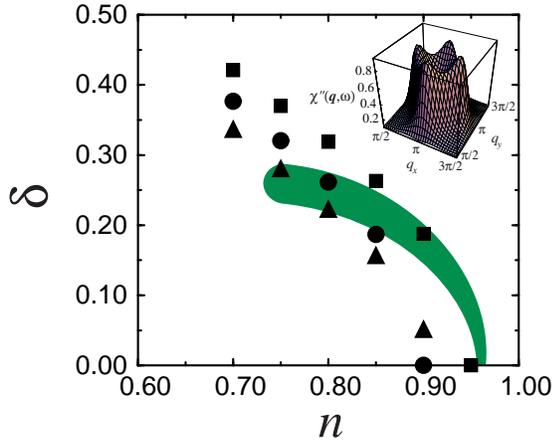}
\caption{
 A plot similar to Fig.\protect\ref{incm-ybco-ncco} for 
$U=4t$, $T=0.05t$, $t''=0.05t$, and $t'=-0.35t$ (solid squares), 
$-0.25t$ (solid circles), $-0.10t$ (solid triangles). 
The hatched area represents the experimental results for LSCO.
\protect\cite{Yamada1} Note $\delta$ defined here is larger
by a factor of 2 than that defined in Ref.\protect\onlinecite{Yamada1}.  
The inset shows a plot similar to Fig.\protect\ref{sus-ybco-ncco} for
$U=4t$, $t'=-0.25t$, $t''=0.05t$, $T=0.05t$,$\omega=0.05t$, and $n=0.80$.
}
\label{incm-lsco}
\end{center}
\end{figure}
\noindent
small values of $-t'$, 
which is inconsistent with experimental observations.
It can be seen that the appearance of 
large $\delta$ with a small amount of hole doping , 
which is observed experimentally ($\delta\sim 0.2$ 
for $x\sim 0.1$),\cite{Yamada1} can only be obtained for large enough $-t'$.
On the other hand, the incommensurability 
increases monotonically with hole doping if we stick 
to a fixed $(t',t'')$. 
Thus we conclude that 
we have to have $-t'/t$ (and possibly $t''/t$) decreasing with hole doping 
in order to obtain the experimentally observed saturation of 
$\delta$.\cite{Yamada1} 
The doping dependence of both the Fermi surface and the spin structure
can then be understood within a single scenario.

To conclude, the main message of the present paper is 
that the experimental observations such as 
the saturation of the incommensurability in LSCO and 
the sharp contrast of the spin structure 
between electron doped and hole doped systems
can be understood within a {\it single-band} model, 
and that we do not have to go back, as one might suspect, 
to the original three-band model that explicitly takes into account 
Cu $3d$ and O $2p$ orbitals.

A remaining problem is why, if our picture is correct, 
$-t'/t$ (and possibly $t''/t$) decreases with hole doping in LSCO. 
As for the discrepancy between the ARPES result and the 
band calculation in the underdoped regime, 
we may make a following conjecture at the present stage.
If we closely look at the band calculation for 
LSCO,\cite{Freeman,Hybertsen2} there is, 
in contrast to that of YBCO,\cite{Pickett} 
a large dispersion between $\Gamma$ (${\bf k}=(0,0)$)
and G$_1$ $({\bf k}=(\pi,0))$ 
in the band lying near the Fermi level.  
In a tight-binding model, this corresponds to a small $-t'$. 
A large energy difference between $\Gamma$ and G$_1$ 
is known to be due to the mixing between the 
Cu $d_{x^2-y^2}$- O $p_{x,y}$ orbital 
and the Cu $d_{3z^2-r^2}$ - O $p_{z}$ orbital.\cite{Freeman}
If, for some reason, this hybridization is 
weaker than expected in the underdoped regime, 
so that the main character of the band near the Fermi level is 
Cu $d_{x^2-y^2}$ - O $p_{x,y}$, then the 
energy difference between $\Gamma$-G$_1$ would be smaller,
amounting to a larger $-t'$.
If this is correct, the reason why the hybridization is 
weak in the underdoped regime, or why it is recovered in the overdoped 
regime becomes a future problem. 
Experimental observation in Ref.\onlinecite{Chen} showing that 
the Cu $d_{3z^2-r^2}$ - O $p_{z}$ character grows with hole doping 
may have some relevance to this point.

The relation between the present result and the so-called `1/8-problem'
or `charge stripes'\cite{Tranquada} is another interesting future problem.

We are indebted to Prof. K. Yamada 
and Prof. Y. Endoh for discussion 
prior to publication of Ref.\onlinecite{Yamada2}.  
We also thank Prof. A. Fujimori
for discussion prior to publication of Ref.\onlinecite{Ino} and 
for pointing out the experimental observation in Ref.\onlinecite{Chen}.
We also thank Dr. H. Kontani for sending us Ref.\onlinecite{Kontani} 
prior to publication.
Numerical calculations were performed at the Supercomputer Center,
ISSP, University of Tokyo.
K.K. acknowledges support by the Grant-in-Aid for Scientific
Research from the Ministry of Education of Japan.
R.A. acknowledges support by the JSPS Research Fellowships for
Young Scientists.

%%%%%%%%%%%%%%%%%%%%  References %%%%%%%%%%%%%%%%%%%%%%%%%%%%%%%%

\end{multicols}

\begin{references}
\bibitem{comment1} The FLEX approximation may not be valid in the 
very vicinity of half-filling, where Mott's transition should occur. 
We nevertheless applied the method 
to obtain a commensurate peak at $n=1.0$, although the peak structure 
turns out to be substantially broad.
\bibitem{comment2}This turns out to be the case for other values of $(t',t'')$
we studied, e.g., $t'=-0.2t$, $t''=0.0$.
\bibitem{comment3} Conversely, if spin structures qualitatively 
different from that of YBCO are to be present in other materials with similar
Fermi surfaces, a smaller $-t'$ with a larger $t''$ could be the cause.
\end{references}
\end{document}